\documentclass[aps,prl,twocolumn,superscriptaddress,showpacs,showkeys]{revtex4-1}
\usepackage{graphicx}
\usepackage{amsmath}
\begin{document}

%\preprint{}

\title{Graphite Under Compression: Shift of Layer Breathing and Shear Modes Frequencies with Interlayer Spacing}


\author{Y. W. Sun}
\email[Corresponding author: ]{yiwei.sun@qmul.ac.uk}
\affiliation{College of Information Science and Electronic Engineering, Zhejiang University, Hangzhou 310027, China}
\author{D. Holec}
\email[Corresponding author (for more calculation details): ]{david.holec@unileoben.ac.at}
\affiliation{Department of Physical Metallurgy and Materials Testing, Montanuniversit\"{a}t Leoben, Leoben 8700, Austria}
\author{Y. Xu}
\affiliation{College of Information Science and Electronic Engineering, Zhejiang University, Hangzhou 310027, China}
\author{D. J. Dunstan}
\affiliation{School of Physics and Astronomy, Queen Mary University of London, London E1 4NS, United Kingdom}

%\thanks{}


\date{\today}

\begin{abstract}
Layered materials have huge potential in various applications due to their extraordinary properties. To determine the interlayer interaction (or equivalently the layer spacing under different perturbations) is of critical importance. In this letter, we focus on one of the most prominent layered materials, graphite, and theoretically quantify the relationship between its interlayer spacing and the vibrational frequencies of its layer breathing and shear modes, which are measures of the interlayer interaction. The method used here to determine the interlayer interaction can be further applied to other layered materials. 
\end{abstract}

%\keywords{}

\maketitle

Layered materials have unique electronic and excellent mechanical properties \cite{Wilson69}. Many of the properties are closely related to the weak interlayer van der Waals (vdW) interaction between the 2-D molecular monolayers \cite{Wilson69}. Vibrational modes include shear modes (CMs) and layer breathing modes (LBMs), due to relative motions of the planes. The CMs (vibrations parallel to the planes) have been experimently identified in many bulk layered materials, such as h-BN \cite{Nemanich81}, MoS$_{2}$ \cite{Verble72} and WSe$_{2}$ \cite{Mead77}. The LBMs (vibrations perpendicular to the planes) are less studied, because they are optically inactive. Nevertheless, both modes are of significant importance to the understanding of various layered materials (and the full exploitation of their application potential), as direct measures of the interlayer interactions. 

The interlayer interactions (and the frequencies of the CMs and LBMs) of layered materials strongly depend on the interlayer spacing. So do many of their properties. Therefore, the control of the interlayer spacing (strain) is vital to the applications -- one could adjust the properties as needed by tuning the interlayer spacing. To quantify the relationship between the frequencies of the interlayer modes and the interlayer strain provides a practically convenient way to determine the strain as the phonons are usually easier to directly detect than the lattice constant in experiments. Also this relationship itself is a fundamental mechanical property for each material.

Among layered materials, we are most familiar with graphite, with just four carbon atoms in a unit cell. Graphite has been studied for over two thousand years. The relatively recent successful synthesis of graphene \cite{Novoselov04}, along with the massive research that followed brought the understanding of graphite to a new level. Yet much remains behind a veil. 

The CM of graphite is a Raman active E$_{2g}$ mode and was measured at ~42 cm$^{-1}$ in 1975 \cite{Nemanich75}. Tan \textit{et al.} measured the CMs of two- to eight-layer graphene and bulk graphite. These frequencies fitted well with a linear chain model \cite{Tan12} :
\begin{equation}
\omega_{N}=\frac{1}{\sqrt{2}\pi c}\sqrt{\frac{\alpha_{CM}}{\mu}}\sqrt{1+\cos(\frac{\pi}{N})},
\label{eqlc}
\end{equation}
where $\omega_{N}$ is in cm$^{-1}$, \textit{N} is the number of layers, \textit{c} is the speed of light in cm s$^{-1}$, $\alpha_{CM}\sim 12.8\times 10^{18}$ Nm$^{-3}$ is the only fitting parameter, referring to the interlayer force constant for the CMs, and $\mu=7.6\times 10^{-27}$ kg\AA{}$^{-2}$ is the mass per unit area of the single-layer graphene. For graphite, $N\rightarrow\infty$ and $\omega_\infty=\frac{1}{\pi c}\sqrt{\frac{\alpha_{CM}}{\mu}}$. The shift of the CM with pressure was measured by Hanfland $\textit{et al.}$ \cite{Hanfland89}. They measured the frequency at ambient pressure at 44 cm$^{-1}$ and fitted their data under pressure with the following equation \cite{Hanfland89}:
\begin{equation}
\omega (P)/\omega_{0}=[(\delta_{0}/\delta^{\prime})P+1]^{\delta^{\prime}},
\label{ME}
\end{equation}
where $\delta_0$ is the logarithmic pressure derivative (d ln$\omega$/d$P$)$_{P=0}$ and $\delta^{\prime}$ is the pressure derivative of d ln$\omega$/d$P$. They obtained $\delta_{0}$=0.110 GPa$^{-1}$ and $\delta^{\prime}$=0.43. 

The pressure-induced change of the CM frequency is intrinsically due to the change of the lattice constants, mainly $a_{33}$. The Gr\"{u}neisen parameter is commonly used to relate phonon frequencies to volume \cite{Weinstein84}. For layered materials of large anisotropy, such as graphite, the interlayer modes and intralayer modes are independently related to out-of-plane strain and in-plane strain, respectively, by a scaling parameter $\gamma$ \cite{Zallen74}:
\begin{equation}
\omega(P)/\omega_{0}=[a(P)/a_{0}]^{-3\gamma}.
\label{gamma3}
\end{equation}
There is one scaling parameter for each mode. The scaling parameter and the Gr\"{u}neisen parameter are equivalent for 3-D isotropic materials. For the CM, Hanfland $\textit{et al.}$ gave ${\gamma}$=1.4, for the pressure range up to 14 GPa \cite{Hanfland89}.

The LBM of graphite is an optically inactive B$_{2g}$ mode and was measured at 127 cm$^{-1}$ by inelastic neutron scattering \cite{Alzyab88}. The LBM of multilayer Bernal-stacked \cite{Bernal24} graphene was measured by combination Raman (LO+ZO$^{\prime}$) \cite{Lui12} and has not been directly probed so far at room temperature due to the low electron-phonon coupling and its symmetry. Wu $\textit{et al.}$ directly measured the LBM of twisted multilayer graphene \cite{Wu15} and similarly to the CM, fitted the position with the linear chain model \cite{Ferrari13}:
\begin{equation}
\omega_{LBM_{NN-i}}=\omega_{LBM_{\infty}}\sin(\frac{i\pi}{2N})
\label{LBM}
\end{equation}
where \textit{N} is the number of layers. 

The shift of the LBM with pressure was measured by Alzyab $\textit{et al.}$ (the same neutron scattering experiment as mentioned above) for the pressure range up to 2 GPa and a shift rate of $\omega^{\prime}$=19 cm$^{-1}$GPa$^{-1}$ was observed \cite{Alzyab88}. From the shift rate, $\delta_{0}$= ($\omega^{\prime}$/$\omega_0$)=0.15 GPa$^{-1}$ were obtained. No further study on the Gr\"{u}neisen parameter of the LBM of graphite or multi-layer graphene has been reported. 

In this paper, we modeled graphite under hydrostatic compression and uniaxial compression along c-axis. We calculated the phonon frequencies of the CMs and LBMs in each case and obtained the Gr\"{u}neisen parameters for each mode.

We used density functional theory (DFT) \cite{DFT1,DFT2} as implemented in the Vienna ab initio Simulation Package (VASP) \cite{VASP} with the generalized gradient approximation as parameterized by Perdew, Burke and Ernzerhof \cite{GGA}. We used 900 eV plane-wave cutoff energy and the reciprocal unit cell meshed with 18$\times$18$\times$9 k-points. We included the vdW using the Grimme method \cite{vdW} as implemented in the VASP code. We used the 2$\times$2$\times$2 supercell employing the finite displacement method as implemented in the PHONOPY code \cite{phonon} to calculate the phonon frequencies. The calculations setup were the same as that used in our previous work \cite{Sun15}, where more details can be found.

We started by obtaining the optimized geometry of unstrained graphite, with the in-plane bond length $a_0$=1.42 \AA{} and the interlayer distance $c_0$=3.20 \AA{}. The errors relative to the experimental values are 0.06\% and 4.6\%, respectively \cite{Bosak07}. We expect a relatively large error in the interlayer spacing, despite the vdW add-on being used. To minimize the effect of the inaccuracy of the vdW, we focused on graphite under compressive strain, where the repulsion dominates over the vdW attraction. Furthermore, the shift of the phonon frequencies with compressive strain is mainly due to the increasing overlap of $\pi$-electrons of neighboring layers, where vdW plays only a small role. 

\begin{figure}
\includegraphics[width=1.1\linewidth]{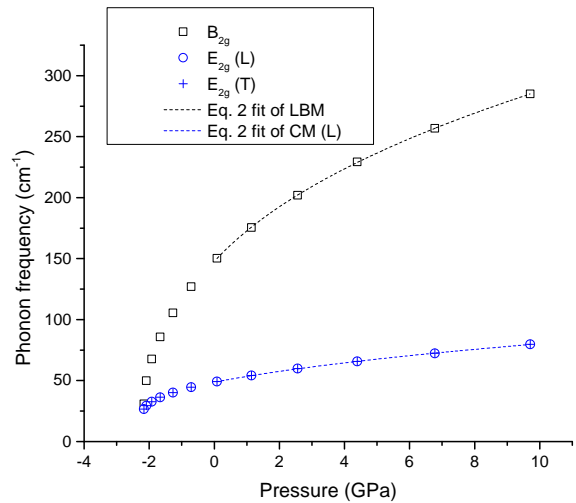}
\caption{(colour online) The frequencies of interlayer modes CM and LBM of graphite are plotted against applied hydrostatic pressure. The fits using Eq. \ref{ME} under compressive pressure, range up to 10 GPa are presented in dash lines, black for the LBM and blue for the CM.}
\label{FP}
\end{figure}

We then modeled graphite under hydrostatic pressure, following the same method as in Ref. \cite{Sun15} -- setting a unit cell volume and calculating the corresponding pressure. The frequencies of the CM (E$_{2g}^{(1)}$ --- (1) is used to distinguish the CM from the intralayer GM \cite{Tuinstra70} of graphite) and LBM (B$_{2g}$) of unstrained graphite are 48.70 and 147.72 cm$^{-1}$, respectively. The errors relative to the experimental values are 10.7\% and 16.3\%, respectively. We now plot the phonon frequencies with pressure in Fig. \ref{FP}.
(L) and (T) refer to the longitudinal and transverse modes, respectively -- two orthogonal in-plane vibrations. From Fig. \ref{FP}, the difference between the CM (L) and CM (T) is merely to be seen and therefore we study the longitudinal mode alone as a representative for the CM. We fit the data under compressive pressure range up to 10 GPa with Eq. \ref{ME} and obtained $\delta_{0}$=0.1055 GPa$^{-1}$, $\delta^{\prime}$=0.3707 for the CM and $\delta_{0}$=0.1969 GPa$^{-1}$, $\delta^{\prime}$=0.3541 for the LBM. As mentioned before, the values from the experiments are $\delta_{0}$=0.110(8) GPa$^{-1}$ and $\delta^{\prime}$=0.43(3) for the CM (pressure range up to 14 GPa) \cite{Hanfland89}, and $\delta_{0}$=0.15 GPa$^{-1}$ with no error bars for the LBM (pressure range up to 2 GPa) \cite{Alzyab88}. For the CM, our results agree reasonably well with the experimental values, especially in the initial shift rate. For the LBM, we expect the calculations to be as reliable as the CM, as the mechanism of the shifts of both modes under pressure is the same (increasing interlayer interaction through overlap of $\pi$-electrons of neighboring layers). The less satisfying agreement between the calculations of the LBM and experiment is probably due to the lack of reliable experimental data.

\begin{figure}
\includegraphics[width=1.1\linewidth]{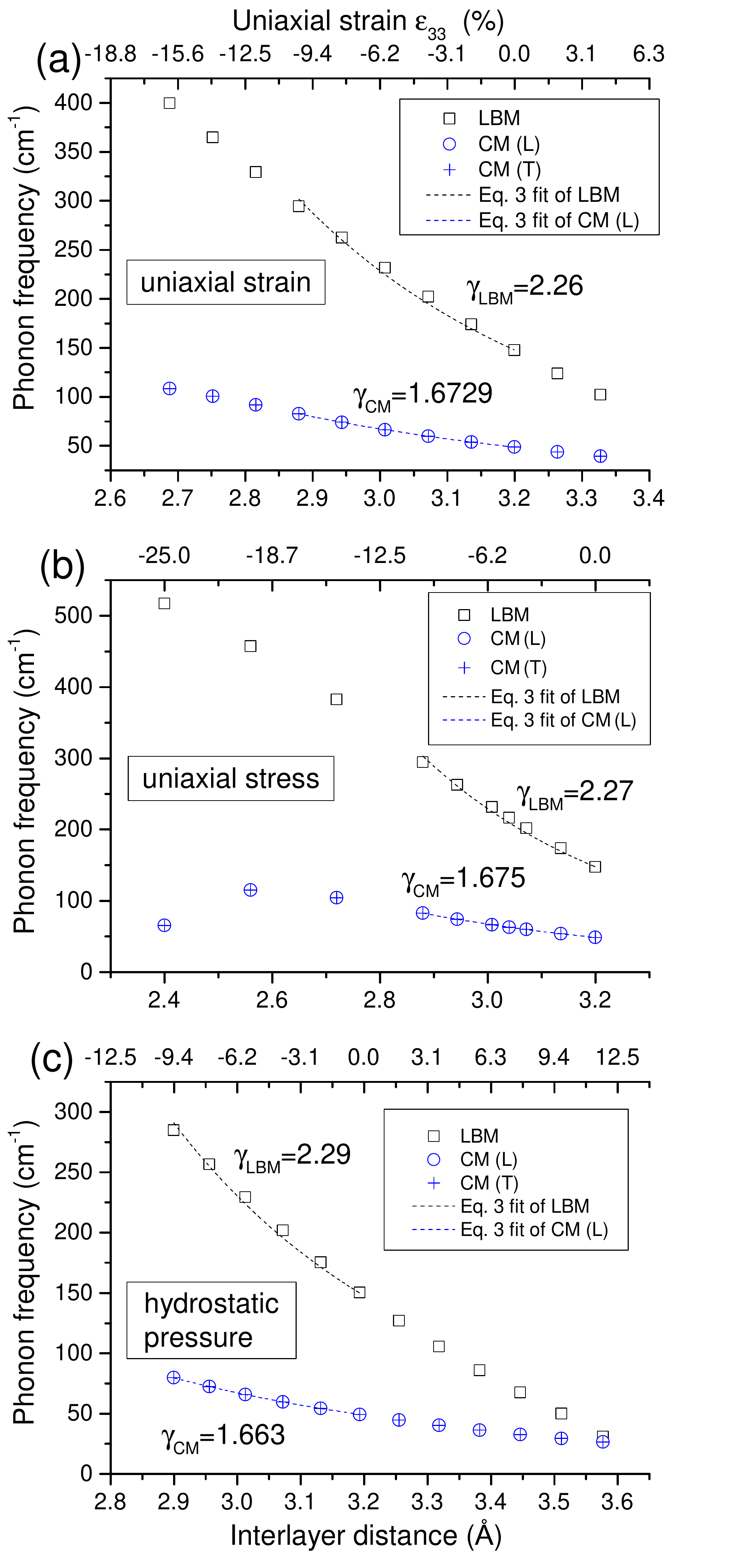}
\caption{(colour online) The frequencies of interlayer modes CM and LBM are plotted against interlayer distance (lower abscissae) and the corresponding strain $\varepsilon_{33}$ (upper abscissae) for (a) uniaxial strain (in-plane geometry fixed), (b) uniaxial stress (in-plane geometry released), and (c) hydrostatic pressure. The fits using Eq. \ref{gamma3} are plotted as dash lines, black for the LBM and blue for the CM. They are plotted over the range 0--10 GPa in (c) and the corresponding range of $\varepsilon_{33}$ in (a) and (b).}
\label{FOS}
\end{figure}

The shift of the frequencies of interlayer modes, such as CM and LBM, should be mainly induced by interlayer strain, but we do not know if there is contribution from intra-layer strain. As our previous work showed, the contribution of interlayer strain to intra-layer modes of graphite was non-negligible \cite{Sun15}. To check, we quantify the relationship between the phonon frequencies and non-hydrostatic strain. We modeled graphite under compression along \textit{c}-axis, by applying uniaxial strain -- setting an interlayer distance and fixing the in-plane geometry, and by applying uniaxial stress -- releasing the in-plane geometry. We present the data under uniaxial strain first. We plot the phonon frequencies with interlayer distance in Fig. \ref{FOS} (a).
We fit the data with Eq. \ref{gamma3} and obtain $\gamma_{CM}$=1.6729 (2) and $\gamma_{LBM}$=2.26 (5). The errors are from the fitting. For uniaxial stress, similarly we fit the data with Eq. \ref{gamma3} and obtain $\gamma_{CM}$=1.675 (1) and $\gamma_{LBM}$=2.27 (5) in Fig. \ref{FOS} (b). The almost identical values of $\gamma_{CM}$ and $\gamma_{LBM}$ from uniaxial strain and stress indicate, that the contributions of the in-plane strain to the CM and LBM are trivial, unlike the non-negligible contributions of the out-of-plane strain to the in-plane GM \cite{Sun15}.

We presented the data under hydrostatic pressure at beginning as they were compared to the experimental results to show the reliability of our calculations. Those data can also be used to further validate the trivial contribution of the in-plane strain to the interlayer modes by obtaining the $\gamma$ under hydrostatic pressure and see if the value is close to the cases of uniaxial strain and stress. We calculated the corresponding interlayer distance to the data in Fig. \ref{FP} (the calculation input is the unit cell volume here), and plot the phonon frequencies against it in Fig. \ref{FOS} (c). We obtain $\gamma_{CM}$=1.663 (2) and $\gamma_{LBM}$=2.29 (4). We are now confident to conclude that for the frequencies of the CM and LBM of graphite under strain, there is negligible contribution from the in-plane strain. From the fitting in Fig. \ref{FOS}, Eq. \ref{gamma3} describes the shift of the CM with strain excellently with \textbf{$\gamma_{CM}$=1.6729 (2)} and the LBM reasonably well with \textbf{$\gamma_{LBM}$=2.26 (5)}, over the range up to 10 GPa.

For the GM of graphene, many papers used the phenomenological equation proposed by Thomsen $\textit{et al.}$ to relate the Gr\"{u}neisen parameters obtained in various conditions \cite{Thomsen02}:
\begin{equation}
-\frac{\Delta \omega}{\omega_0}=\gamma^{\prime}(\varepsilon_{xx}+\varepsilon_{yy})\pm\frac{1}{2}SDP(\varepsilon_{xx}-\varepsilon_{yy}),
\label{Tom}
\end{equation} 
where $\omega_0$ is the unperturbed GM frequency and the $\textit{SDP}$ is the shear deformation potential. We use $\gamma^{\prime}$ here to distinguish from the previous scaling parameter $\gamma$. Eq. \ref{Tom} makes explicit the two-dimensional nature of the analysis and later Huang \textit{et al.} derived this equation from the dynamical equation \cite{Huang09}:
\begin{equation}
\begin{pmatrix} \omega_0^2+A\varepsilon_{xx}+B\varepsilon_{yy} & 0 \\ 0 & \omega_0^2+B\varepsilon_{xx}+A\varepsilon_{yy} \end{pmatrix} \begin{pmatrix} u_1 \\ u_2 \end{pmatrix} = \omega^2 \\ \begin{pmatrix} u_1 \\ u_2 \end{pmatrix},
\label{de}
\end{equation}
where \textbf{u}=($u_1$,$u_2$) is the relative displacement of the two carbon atoms in the unit cell, and \textit{A} and \textit{B} are two independent parameters from the hexagonal symmetry. $\gamma^{\prime}=-(A+B)/4\omega_{0}^2$ and $SDP=-(B-A)/2\omega_0^2$. We follow this method to see if it describes the shifts of the CMs and LBMs of graphite better than Eq. \ref{gamma3}. Since the in-plane strain makes negligible contribution to the two interlayer modes of graphite, the analysis becomes one-dimensional:
\begin{equation}
(\omega_0^2-C\varepsilon_{zz})u=\omega^2u,
\label{ph}
\end{equation}
where \textit{u} is the displacement of the layers. We fit the data in Fig. \ref{FOS} with this Eq. \ref{ph} and obtain
\begin{figure}
	\includegraphics[width=1.1\linewidth]{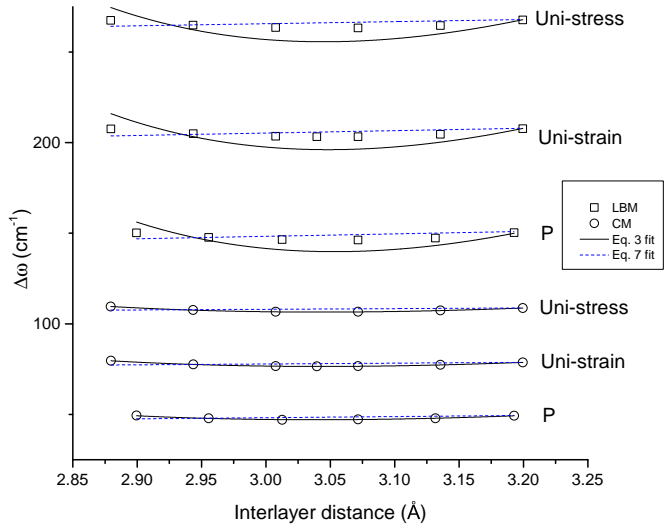}
	\caption{(colour online) Straight lines are subtracted from all the data in Fig. \ref{FOS} to make the data roughly on horizontal lines. Then they are vertically shifted for clarity and labeled. The fits using Eq. \ref{gamma3} are plotted as black solid lines and the fits using Eq. \ref{ph} are plotted as blue dash lines.}
	\label{FOSs}
\end{figure}
$C_{CM}$=3.9 (2)$\times$10$^{4}$, $C_{LBM}$=5.8 (3)$\times$10$^{5}$ under uniaxial strain, $C_{CM}$=3.8 (2)$\times$10$^{4}$, $C_{LBM}$=5.7 (3)$\times$10$^{5}$ under uniaxial stress, and $C_{CM}$=3.7 (2)$\times$10$^{4}$, $C_{LBM}$=5.7 (3)$\times$10$^{5}$ under hydrostatic pressure. The unit of \textit{C} is cm$^{-2}$. More sensibly, we obtain the $\gamma^{\prime}$ in a similar form to that in Eq. \ref{Tom}:
\begin{equation}
-\frac{\Delta \omega}{\omega_0}=\gamma^{\prime}\varepsilon_{zz},
\end{equation}
where $\gamma^{\prime}$=$C/(\omega_{0}(\omega+\omega_{0})$). A caveat must be stated. For the GM, the frequency shift is fractionally small and therefore $\omega+\omega_{0}\approx 2\omega_{0}$, giving a linear relationship between the frequency and strain. To compare the fittings with Eq. \ref{ph} to Eq. \ref{gamma3}, we subtract straight lines from all the data in Fig. \ref{FOS} to make the data roughly on horizontal lines, and then vertically shift them for clarity. We plot the fitting curves of all the data with both Eq. \ref{gamma3} and Eq. \ref{ph} in Fig. \ref{FOSs}. Here the shift rates of the frequencies of the interlayer modes fitted with Eq. \ref{ph} drop with increasing strain, opposite to the fitting with Eq. \ref{gamma3}. We obtain $\gamma^{\prime}_{CM}=6.1$ and $\gamma^{\prime}_{LBM}=8.9$ under uniaxial strain, over the range up to 10 GPa but clearly this fitting following the phenomenological method is not as good as that with Eq. \ref{gamma3} for the CM. For the LBM, the fitting with Eq. \ref{gamma3} agrees with the data on superlinearity, while Eq. \ref{ph} shows sublinearity. Therefore, we think that $\gamma$ describes the shift of interlayer modes with strain better than $\gamma^{\prime}$.

Finally, to extend this determination method of interlayer spacing from bulk to multilayer materials, we need to relate the $\gamma$ to the interlayer coupling strength $\alpha$ in Eq. \ref{eqlc}. We do not consider the change in the mass per unit area of single-layer graphene $\mu$ as the in-plane contribution is negligible. So it is
\begin{equation}
\alpha=\alpha_{0}(a/a_0)^{6\gamma},
\end{equation}
both for the CM and the LBMs.

In conclusion, we modeled graphite under hydrostatic pressure and uniaxial compression along \textit{c}-axis. we calculated the phonon frequencies of the CM and LBM at various interlayer distance. We quantified the relationship between these two, separately by $\gamma$ derived from the original definition of the Gr\"{u}neisen parameter and by $\gamma^{\prime}$, following a phenomenological method. We found that the former method describes the relationship better and the in-plane strain makes a negligible contribution to the shifts of the interlayer modes. The $\gamma$ can be further related to the interlayer coupling strength in the linear chain model and Eq. \ref{eqlc} and \ref{LBM} can now be used to determine the interlayer strain in multilayer graphene. This strain determination method can also be applied to other layer materials in both bulk and multilayer forms. 

\section{}
\subsection{}
\subsubsection{}

\bibliography{apssamp}

\begin{thebibliography}{26}%
\makeatletter
\providecommand \@ifxundefined [1]{%
 \@ifx{#1\undefined}
}%
\providecommand \@ifnum [1]{%
 \ifnum #1\expandafter \@firstoftwo
 \else \expandafter \@secondoftwo
 \fi
}%
\providecommand \@ifx [1]{%
 \ifx #1\expandafter \@firstoftwo
 \else \expandafter \@secondoftwo
 \fi
}%
\providecommand \natexlab [1]{#1}%
\providecommand \enquote  [1]{``#1''}%
\providecommand \bibnamefont  [1]{#1}%
\providecommand \bibfnamefont [1]{#1}%
\providecommand \citenamefont [1]{#1}%
\providecommand \href@noop [0]{\@secondoftwo}%
\providecommand \href [0]{\begingroup \@sanitize@url \@href}%
\providecommand \@href[1]{\@@startlink{#1}\@@href}%
\providecommand \@@href[1]{\endgroup#1\@@endlink}%
\providecommand \@sanitize@url [0]{\catcode `\\12\catcode `\$12\catcode
  `\&12\catcode `\#12\catcode `\^12\catcode `\_12\catcode `\%12\relax}%
\providecommand \@@startlink[1]{}%
\providecommand \@@endlink[0]{}%
\providecommand \url  [0]{\begingroup\@sanitize@url \@url }%
\providecommand \@url [1]{\endgroup\@href {#1}{\urlprefix }}%
\providecommand \urlprefix  [0]{URL }%
\providecommand \Eprint [0]{\href }%
\providecommand \doibase [0]{http://dx.doi.org/}%
\providecommand \selectlanguage [0]{\@gobble}%
\providecommand \bibinfo  [0]{\@secondoftwo}%
\providecommand \bibfield  [0]{\@secondoftwo}%
\providecommand \translation [1]{[#1]}%
\providecommand \BibitemOpen [0]{}%
\providecommand \bibitemStop [0]{}%
\providecommand \bibitemNoStop [0]{.\EOS\space}%
\providecommand \EOS [0]{\spacefactor3000\relax}%
\providecommand \BibitemShut  [1]{\csname bibitem#1\endcsname}%
\let\auto@bib@innerbib\@empty
%</preamble>
\bibitem [{\citenamefont {Wilson}\ \emph {et~al.}(1969)\citenamefont {Wilson}
  \emph {et~al.}}]{Wilson69}%
  \BibitemOpen
  \bibfield  {author} {\bibinfo {author} {\bibfnamefont {J.~A.}\ \bibnamefont
  {Wilson}} \emph {et~al.},\ }\href@noop {} {\bibfield  {journal} {\bibinfo
  {journal} {Adv. Phys.}\ }\textbf {\bibinfo {volume} {18}},\ \bibinfo {pages}
  {193} (\bibinfo {year} {1969})}\BibitemShut {NoStop}%
\bibitem [{\citenamefont {Nemanich}\ \emph {et~al.}(1981)\citenamefont
  {Nemanich} \emph {et~al.}}]{Nemanich81}%
  \BibitemOpen
  \bibfield  {author} {\bibinfo {author} {\bibfnamefont {R.~J.}\ \bibnamefont
  {Nemanich}} \emph {et~al.},\ }\href@noop {} {\bibfield  {journal} {\bibinfo
  {journal} {Phys. Rev. B}\ }\textbf {\bibinfo {volume} {23}},\ \bibinfo
  {pages} {6348} (\bibinfo {year} {1981})}\BibitemShut {NoStop}%
\bibitem [{\citenamefont {Verble}\ \emph {et~al.}(1972)\citenamefont {Verble}
  \emph {et~al.}}]{Verble72}%
  \BibitemOpen
  \bibfield  {author} {\bibinfo {author} {\bibfnamefont {J.~L.}\ \bibnamefont
  {Verble}} \emph {et~al.},\ }\href@noop {} {\bibfield  {journal} {\bibinfo
  {journal} {Solid State Commun.}\ }\textbf {\bibinfo {volume} {11}},\ \bibinfo
  {pages} {941} (\bibinfo {year} {1972})}\BibitemShut {NoStop}%
\bibitem [{\citenamefont {Mead}\ \emph {et~al.}(1977)\citenamefont {Mead} \emph
  {et~al.}}]{Mead77}%
  \BibitemOpen
  \bibfield  {author} {\bibinfo {author} {\bibfnamefont {D.~G.}\ \bibnamefont
  {Mead}} \emph {et~al.},\ }\href@noop {} {\bibfield  {journal} {\bibinfo
  {journal} {Can. J. Phys.}\ }\textbf {\bibinfo {volume} {55}},\ \bibinfo
  {pages} {379} (\bibinfo {year} {1977})}\BibitemShut {NoStop}%
\bibitem [{\citenamefont {Novoselov}\ \emph {et~al.}(2004)\citenamefont
  {Novoselov} \emph {et~al.}}]{Novoselov04}%
  \BibitemOpen
  \bibfield  {author} {\bibinfo {author} {\bibfnamefont {K.~S.}\ \bibnamefont
  {Novoselov}} \emph {et~al.},\ }\href@noop {} {\bibfield  {journal} {\bibinfo
  {journal} {Science}\ }\textbf {\bibinfo {volume} {306}},\ \bibinfo {pages}
  {666} (\bibinfo {year} {2004})}\BibitemShut {NoStop}%
\bibitem [{\citenamefont {Nemanich}\ \emph {et~al.}(1975)\citenamefont
  {Nemanich} \emph {et~al.}}]{Nemanich75}%
  \BibitemOpen
  \bibfield  {author} {\bibinfo {author} {\bibfnamefont {R.~J.}\ \bibnamefont
  {Nemanich}} \emph {et~al.},\ }in\ \href@noop {} {\emph {\bibinfo {booktitle}
  {Proceedings of the International Conference on Lattice Dynamics}}},\
  \bibinfo {editor} {edited by\ \bibinfo {editor} {\bibfnamefont
  {M.}~\bibnamefont {Balkanski}}}\ (\bibinfo  {publisher} {Flammarion},\
  \bibinfo {address} {Paris},\ \bibinfo {year} {1975})\BibitemShut {NoStop}%
\bibitem [{\citenamefont {Tan}\ \emph {et~al.}(2012)\citenamefont {Tan} \emph
  {et~al.}}]{Tan12}%
  \BibitemOpen
  \bibfield  {author} {\bibinfo {author} {\bibfnamefont {P.~H.}\ \bibnamefont
  {Tan}} \emph {et~al.},\ }\href@noop {} {\bibfield  {journal} {\bibinfo
  {journal} {Nat. Mater.}\ }\textbf {\bibinfo {volume} {11}},\ \bibinfo {pages}
  {294} (\bibinfo {year} {2012})}\BibitemShut {NoStop}%
\bibitem [{\citenamefont {Hanfland}\ \emph {et~al.}(1989)\citenamefont
  {Hanfland} \emph {et~al.}}]{Hanfland89}%
  \BibitemOpen
  \bibfield  {author} {\bibinfo {author} {\bibfnamefont {M.}~\bibnamefont
  {Hanfland}} \emph {et~al.},\ }\href@noop {} {\bibfield  {journal} {\bibinfo
  {journal} {Phys. Rev. B}\ }\textbf {\bibinfo {volume} {39}},\ \bibinfo
  {pages} {12598} (\bibinfo {year} {1989})}\BibitemShut {NoStop}%
\bibitem [{\citenamefont {Weinstein}\ \emph {et~al.}(1984)\citenamefont
  {Weinstein} \emph {et~al.}}]{Weinstein84}%
  \BibitemOpen
  \bibfield  {author} {\bibinfo {author} {\bibfnamefont {B.}~\bibnamefont
  {Weinstein}} \emph {et~al.},\ }in\ \href@noop {} {\emph {\bibinfo {booktitle}
  {Light Scattering in Solids}}},\ \bibinfo {editor} {edited by\ \bibinfo
  {editor} {\bibfnamefont {M.}~\bibnamefont {Cardona}} \emph {et~al.}}\
  (\bibinfo  {publisher} {Springer},\ \bibinfo {address} {Berlin},\ \bibinfo
  {year} {1984})\BibitemShut {NoStop}%
\bibitem [{\citenamefont {Zallen}(1974)}]{Zallen74}%
  \BibitemOpen
  \bibfield  {author} {\bibinfo {author} {\bibfnamefont {R.}~\bibnamefont
  {Zallen}},\ }\href@noop {} {\bibfield  {journal} {\bibinfo  {journal} {Phys.
  Rev. B}\ }\textbf {\bibinfo {volume} {9}},\ \bibinfo {pages} {4485} (\bibinfo
  {year} {1974})}\BibitemShut {NoStop}%
\bibitem [{\citenamefont {Alzyab}\ \emph {et~al.}(1988)\citenamefont {Alzyab}
  \emph {et~al.}}]{Alzyab88}%
  \BibitemOpen
  \bibfield  {author} {\bibinfo {author} {\bibfnamefont {B.}~\bibnamefont
  {Alzyab}} \emph {et~al.},\ }\href@noop {} {\bibfield  {journal} {\bibinfo
  {journal} {Phys. Rev. B}\ }\textbf {\bibinfo {volume} {38}},\ \bibinfo
  {pages} {1544} (\bibinfo {year} {1988})}\BibitemShut {NoStop}%
\bibitem [{\citenamefont {Bernal}(1924)}]{Bernal24}%
  \BibitemOpen
  \bibfield  {author} {\bibinfo {author} {\bibfnamefont {J.}~\bibnamefont
  {Bernal}},\ }\href@noop {} {\bibfield  {journal} {\bibinfo  {journal} {Proc.
  R. Soc. Lond. A}\ }\textbf {\bibinfo {volume} {104}},\ \bibinfo {pages} {749}
  (\bibinfo {year} {1924})}\BibitemShut {NoStop}%
\bibitem [{\citenamefont {Lui}\ \emph {et~al.}(2012)\citenamefont {Lui} \emph
  {et~al.}}]{Lui12}%
  \BibitemOpen
  \bibfield  {author} {\bibinfo {author} {\bibfnamefont {C.~H.}\ \bibnamefont
  {Lui}} \emph {et~al.},\ }\href@noop {} {\bibfield  {journal} {\bibinfo
  {journal} {Nano Lett.}\ }\textbf {\bibinfo {volume} {12}},\ \bibinfo {pages}
  {5539} (\bibinfo {year} {2012})}\BibitemShut {NoStop}%
\bibitem [{\citenamefont {Wu}\ \emph {et~al.}(2015)\citenamefont {Wu} \emph
  {et~al.}}]{Wu15}%
  \BibitemOpen
  \bibfield  {author} {\bibinfo {author} {\bibfnamefont {J.}~\bibnamefont {Wu}}
  \emph {et~al.},\ }\href@noop {} {\bibfield  {journal} {\bibinfo  {journal}
  {ACSNano}\ }\textbf {\bibinfo {volume} {9}},\ \bibinfo {pages} {7440}
  (\bibinfo {year} {2015})}\BibitemShut {NoStop}%
\bibitem [{\citenamefont {Ferrari}\ \emph {et~al.}(2013)\citenamefont {Ferrari}
  \emph {et~al.}}]{Ferrari13}%
  \BibitemOpen
  \bibfield  {author} {\bibinfo {author} {\bibfnamefont {A.~C.}\ \bibnamefont
  {Ferrari}} \emph {et~al.},\ }\href@noop {} {\bibfield  {journal} {\bibinfo
  {journal} {Nat. Nanotechnol.}\ }\textbf {\bibinfo {volume} {8}},\ \bibinfo
  {pages} {235} (\bibinfo {year} {2013})}\BibitemShut {NoStop}%
\bibitem [{\citenamefont {Hohenberg}\ \emph {et~al.}(1964)\citenamefont
  {Hohenberg} \emph {et~al.}}]{DFT1}%
  \BibitemOpen
  \bibfield  {author} {\bibinfo {author} {\bibfnamefont {P.}~\bibnamefont
  {Hohenberg}} \emph {et~al.},\ }\href@noop {} {\bibfield  {journal} {\bibinfo
  {journal} {Phys. Rev.}\ }\textbf {\bibinfo {volume} {136}},\ \bibinfo {pages}
  {B864} (\bibinfo {year} {1964})}\BibitemShut {NoStop}%
\bibitem [{\citenamefont {Kohn}\ \emph {et~al.}(1965)\citenamefont {Kohn} \emph
  {et~al.}}]{DFT2}%
  \BibitemOpen
  \bibfield  {author} {\bibinfo {author} {\bibfnamefont {W.}~\bibnamefont
  {Kohn}} \emph {et~al.},\ }\href@noop {} {\bibfield  {journal} {\bibinfo
  {journal} {Phys. Rev.}\ }\textbf {\bibinfo {volume} {140}},\ \bibinfo {pages}
  {A1133} (\bibinfo {year} {1965})}\BibitemShut {NoStop}%
\bibitem [{\citenamefont {Kresse}\ \emph {et~al.}(1996)\citenamefont {Kresse}
  \emph {et~al.}}]{VASP}%
  \BibitemOpen
  \bibfield  {author} {\bibinfo {author} {\bibfnamefont {G.}~\bibnamefont
  {Kresse}} \emph {et~al.},\ }\href@noop {} {\bibfield  {journal} {\bibinfo
  {journal} {Phys. Rev. B}\ }\textbf {\bibinfo {volume} {54}},\ \bibinfo
  {pages} {11169} (\bibinfo {year} {1996})}\BibitemShut {NoStop}%
\bibitem [{\citenamefont {Perdew}\ \emph {et~al.}(1996)\citenamefont {Perdew}
  \emph {et~al.}}]{GGA}%
  \BibitemOpen
  \bibfield  {author} {\bibinfo {author} {\bibfnamefont {J.}~\bibnamefont
  {Perdew}} \emph {et~al.},\ }\href@noop {} {\bibfield  {journal} {\bibinfo
  {journal} {Phys. Rev. Lett.}\ }\textbf {\bibinfo {volume} {77}},\ \bibinfo
  {pages} {3865} (\bibinfo {year} {1996})}\BibitemShut {NoStop}%
\bibitem [{\citenamefont {Grimme}(2006)}]{vdW}%
  \BibitemOpen
  \bibfield  {author} {\bibinfo {author} {\bibfnamefont {S.}~\bibnamefont
  {Grimme}},\ }\href@noop {} {\bibfield  {journal} {\bibinfo  {journal} {J.
  Comput. Chem.}\ }\textbf {\bibinfo {volume} {27}},\ \bibinfo {pages} {1787}
  (\bibinfo {year} {2006})}\BibitemShut {NoStop}%
\bibitem [{\citenamefont {Togo}\ \emph {et~al.}(2008)\citenamefont {Togo} \emph
  {et~al.}}]{phonon}%
  \BibitemOpen
  \bibfield  {author} {\bibinfo {author} {\bibfnamefont {A.}~\bibnamefont
  {Togo}} \emph {et~al.},\ }\href@noop {} {\bibfield  {journal} {\bibinfo
  {journal} {Phys. Rev. B}\ }\textbf {\bibinfo {volume} {78}},\ \bibinfo
  {pages} {134106} (\bibinfo {year} {2008})}\BibitemShut {NoStop}%
\bibitem [{\citenamefont {Sun}\ \emph {et~al.}(2015)\citenamefont {Sun} \emph
  {et~al.}}]{Sun15}%
  \BibitemOpen
  \bibfield  {author} {\bibinfo {author} {\bibfnamefont {Y.~W.}\ \bibnamefont
  {Sun}} \emph {et~al.},\ }\href@noop {} {\bibfield  {journal} {\bibinfo
  {journal} {Phys. Rev. B}\ }\textbf {\bibinfo {volume} {92}},\ \bibinfo
  {pages} {094108} (\bibinfo {year} {2015})}\BibitemShut {NoStop}%
\bibitem [{\citenamefont {Bosak}\ \emph {et~al.}(2007)\citenamefont {Bosak}
  \emph {et~al.}}]{Bosak07}%
  \BibitemOpen
  \bibfield  {author} {\bibinfo {author} {\bibfnamefont {A.}~\bibnamefont
  {Bosak}} \emph {et~al.},\ }\href@noop {} {\bibfield  {journal} {\bibinfo
  {journal} {Phys. Rev. B}\ }\textbf {\bibinfo {volume} {75}},\ \bibinfo
  {pages} {153408} (\bibinfo {year} {2007})}\BibitemShut {NoStop}%
\bibitem [{\citenamefont {Tuinstra}\ \emph {et~al.}(1970)\citenamefont
  {Tuinstra} \emph {et~al.}}]{Tuinstra70}%
  \BibitemOpen
  \bibfield  {author} {\bibinfo {author} {\bibfnamefont {F.}~\bibnamefont
  {Tuinstra}} \emph {et~al.},\ }\href@noop {} {\bibfield  {journal} {\bibinfo
  {journal} {J. Chem. Phys.}\ }\textbf {\bibinfo {volume} {53}},\ \bibinfo
  {pages} {1126} (\bibinfo {year} {1970})}\BibitemShut {NoStop}%
\bibitem [{\citenamefont {Thomsen}\ \emph {et~al.}(2002)\citenamefont {Thomsen}
  \emph {et~al.}}]{Thomsen02}%
  \BibitemOpen
  \bibfield  {author} {\bibinfo {author} {\bibfnamefont {C.}~\bibnamefont
  {Thomsen}} \emph {et~al.},\ }\href@noop {} {\bibfield  {journal} {\bibinfo
  {journal} {Phys. Rev. B}\ }\textbf {\bibinfo {volume} {65}},\ \bibinfo
  {pages} {073403} (\bibinfo {year} {2002})}\BibitemShut {NoStop}%
\bibitem [{\citenamefont {Huang}\ \emph {et~al.}(2009)\citenamefont {Huang}
  \emph {et~al.}}]{Huang09}%
  \BibitemOpen
  \bibfield  {author} {\bibinfo {author} {\bibfnamefont {M.}~\bibnamefont
  {Huang}} \emph {et~al.},\ }\href@noop {} {\bibfield  {journal} {\bibinfo
  {journal} {Proc. Natl. Acad. Sci.}\ }\textbf {\bibinfo {volume} {106}},\
  \bibinfo {pages} {7304} (\bibinfo {year} {2009})}\BibitemShut {NoStop}%
\end{thebibliography}%

\end{document}